\documentclass[sigconf,nonacm]{acmart}
\usepackage[utf8]{inputenc}

\usepackage[skip=0pt]{caption} 
\usepackage{todonotes}
\usepackage{tikz}
\newcommand*\circled[1]{\tikz[baseline=(char.base)]{
            \node[shape=circle,draw,inner sep=1pt] (char) {#1};}}

\title{Recommending Code Understandability Improvements based on Code Reviews}
\author{Delano Oliveira}
\orcid{0000-0001-6815-9251}
\affiliation{%
   \institution{Federal University of Pernambuco, Recife, Brazil, dho@cin.ufpe.br}
}

\begin{abstract}
Developers spend 70\% of their time understanding code. Code that is easy to read can save time, while hard-to-read code can lead to the introduction of bugs. However, it is difficult to establish what makes code more understandable. Although there are guides and directives on improving code understandability, in some contexts, these practices can have a detrimental effect. Practical software development projects often employ code review to improve code quality, including understandability. Reviewers are often senior developers who have contributed extensively to projects and have an in-depth understanding of the impacts of different solutions on code understandability. This paper is an early research proposal to recommend code understandability improvements based on code reviewer knowledge. The core of the proposal comprises a dataset of code understandability improvements extracted from code reviews. This dataset will serve as a basis to train machine learning systems to recommend understandability improvements.
\end{abstract}

\begin{document}

\maketitle

\section{Introduction}
Code understanding is an essential task in software development. 
Whenever a developer needs to change the code or introduce new  features, it is necessary to understand the previously written code. A study about developer tasks \cite{minelli2015} found out that developers spend 70\% of their time understanding code. This indicates that  easy-to-read code can save time and money spent on software development. However, code that is hard to read can compromise program comprehension and lead to the introduction of bugs.

Researchers have studied  some code patterns, such as code smells~\cite{fowler2018} and atoms of confusion~\cite{Gopstein2017}, that can cause misunderstandings.
To avoid these potentially error-prone code patterns and improve code understandability, software organizations and projects have adopted code conventions and style guides~\cite{smit2011}. Additionally, they use manual (code review) and automatic (static analysis tools) approaches to identify good practices violations (aka. rules violations). On the one hand, code review help to find error-prone code and suggest improvements, but it is a time-consuming activity, and only a minority of the rules violations are flagged and removed~\cite{han2020}. On the other hand, static analysis tools (e.g., Checkstyle) can automatically detect rules violations and fix some of them. However, they report warnings that may or may not correspond to an actual mistake, leading developers to ignore many of those~\cite{marcilio2019static}.

\textit{Observation 1: Existing approaches to improving code understandability are either too expensive to scale up or imprecise.}

Still, even if it was possible to detect and repair all rule violations, there is no guarantee that these conventions improve code understandability. For example, at the time that we write this paper, the Linux kernel coding style\footnote{\url{https://www.kernel.org/doc/html/v4.10/process/coding-style.html}} recommends omitting curly braces following an \texttt{if} statement when the block is a single statement. It has been experimentally verified that this convention may confuse developers~\cite{Gopstein2017}.

Defining which code alternatives, i.e., different code snippets that are functionally equivalent, are easier to read is challenging, and researchers have performed some experiments. Nevertheless, we can find divergent results in some of those studies that prevent us from identifying the best solutions. For example, while Miara et al. \cite{Miara1983} found that two indentation level is better than other alternatives, Bauer et al. \cite{Bauer2019} did not find a statistical difference between levels (0, 2, 4, and 8). Likewise, Medeiros et al. \cite{Medeiros2019} evaluated some atoms of confusion studied by Gopstein et al. \cite{Gopstein2017} and found that 4 of 15 of those patterns (e.g., conditional operator) did not cause misunderstanding. The different contexts considered in experiments can explain different results. In this sense, the study by Oliveira et al. \cite{Oliveira2020} investigated what tasks are performed by subjects and what response variables are employed in those studies to evaluate code readability, as well as what cognitive skills are required by those tasks. In that work, they found various methods to assess readability, and each method will simulate a specific context. Furthermore, they observed that few studies evaluated readability in a way that simulates real-world scenarios.

\textit{Observation 2: There are contradicting results in the literature about which code constructs, styles, and idioms improve understandability.}

\textbf{Problem statement.} Starting from Observations 1 and 2, we can conclude that it is difficult to establish what improves code understandability in general since it also depends on who reads the code (developers). Furthermore, in an ideal world, we would like to combine the expertise involved in code review with the ability to scale of static analysis tools. Therefore, there is a lack of knowledge and tools that recommend code understandability improvements based on the point of view of developers.

This paper is an early research proposal to recommend code understandability improvements based on code reviews.

\section{Proposed Approach}

To tackle this problem, we need to look at how developers improve code understandability in practice, and code reviews are essential to understand that. Code review is a step in the development process aiming to enhance code quality. This practice was primarily motivated by finding defects. However, studies have found that the most often code review outcomes are code improvements. For example, researchers identified that ensuring code readability is an important motivation for code review~\cite{sadowski2018}.

Code review is popular in both industrial and open-source projects. They have adopted this practice with the support of tools. For instance, GitHub is a popular tool that integrates code review in the pull-based model. In that model, developers submit their code changes through pull requests, and reviewers inspect the code, searching for error-prone code and suggesting alternative solutions where they think this is important. Subsequently, developers evaluate those suggestions, apply new changes, and submit them to be revised again. That process repeats until the reviewer judges that the code is good enough to be integrated into the final product. Note that in this process, the reviewer is considered the specialist in code quality, since she is responsible for accepting or not the code changes. Reviewers are often senior developers who have made multiple contributions to a project and understand well what constitutes a good solution.

\textit{Observation 3: Code reviewers are the experts in code quality for their projects.}

Therefore, code review is a process where the developer improves the code quality with the support of a specialist (the reviewer). We are interested in the understandability improvements.

\begin{figure}[t]
    \centering
    \includegraphics[scale=0.55]{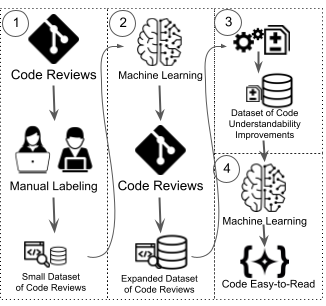}
    \caption{Design of our approach}
    \label{fig:approach}
\end{figure}

The goal of this work is to recommend code understandability improvements to developers. Recent studies \cite{roy2020, tufan2021towards} proposed approaches to detect and recommend code understandability improvements. However, while Tufano et al. \cite{tufan2021towards} considered all sorts of improvements found in code reviews, Roy et al. \cite{roy2020} did not consider the changes triggered during the code review process. In another way, we will build a dataset of code understandability improvements based on the knowledge and experience of code reviewers and use it to train a machine learning system to translate hard-to-read code into a code easier to read. Our approach is presented in \autoref{fig:approach} and is divided into four steps.

In step \circled{1}, we will build a dataset of code reviews related to code understandability. To that, we will classify manually code reviews of selected projects on GitHub. Next, in step \circled{2}, we will expand our dataset. Inspired by the work of Ebert et al. \cite{ebert2017confusion}, we will use our dataset \circled{1} to train a machine learning system to classify more code reviews. The focus, in this case, is on achieving high precision, even if that results in a relatively low recall. So, in step \circled{3}, we will build a dataset of code understandability improvements. In that sense, we will extract code changes triggered by code reviews classified as about understandability in our expanded dataset \circled{2}. The dataset of step \circled{3} will contain pairs ($v, v+1$), where $v$ is the version before the code understandability improvement, and $v+1$ is the version after that improvement. Finally, in step \circled{4}, we will use our dataset of code understandability improvements \circled{3} as input to train a machine learning system to translate code hard-to-read into code easy-to-read.

At present, we are working on the manual classification of code reviews, where our goal is to evaluate if the comments of the reviewer intend to improve understandability. For example, the comment\footnote{\url{https://github.com/apache/beam/pull/13435\#discussion_r542297878}} (``\textit{if \{\} else \{\} or return fieldDescriptor.requirementType == REQUIRED ? ... : ...} would be neater as there is only one case'') suggests that \textit{if and else} or \textit{operator ternary} would be more understandable than a \textit{switch} (solution proposed by the developer). In this case, we assume that the code neater intends to make it easier to read. However, sometimes it is necessary to look beyond the text. For instance, we found a comment\footnote{\url{https://github.com/apache/skywalking/pull/4332\#discussion_r376758126}} (``This could be replaced by simplier foreach.''), where the suggestion of the reviewer can indicate a solution only less verbose. However, the code shows us that the intent is to make the code more verbose to improve understandability.



\section{Conclusion Remarks}

Code review is a widely adopted practice to improve code quality. Moreover, with a large amount of data available, code review is a rich research source to approach based on machine learning. Despite this, we have little knowledge of how to automatically improve code quality based on code reviews. Recent research \cite{tufan2021towards, Hellendoorn2021} has taken the first steps in this direction. However, these are still initial steps and more research is needed to make these models usable by developers. Following along these lines, this paper proposes an approach that uses a machine learning system to improve code understandability based on code reviews.

\noindent\textbf{Acknowledgements.} We thank the anonymous reviewers for their valuable feedback on this paper. This research was partially funded by CAPES/Brazil (88887.582851/2020-0) and the Swedish Foundation for Strategic Research under the TrustFull project.

\vspace{-7pt}
\bibliography{bibliography}

\end{document}